# Ultrasensitive barocaloric material for room-temperature solid-state refrigeration

Qingyong Ren[1,2,#], Ji Qi[3,4,#], Dehong Yu[5], Wenli Song[1,2], Bao Yuan[1,2], Tianhao Wang[1,2], Weijun Ren,[3] Zhidong Zhang[3,4], Xin Tong[1,2],* & Bing Li[3,4],*

[1]Institute of High Energy Physics, Chinese Academy of Sciences (CAS), Beijing 100049, China
[2]Spallation Neutron Source Science Center, Dongguan 523803, China
[3]Shenyang National Laboratory for Materials Science, Institute of Metal Research, Chinese Academy of Sciences, 72 Wenhua Road, Shenyang, Liaoning 110016, China.
[4]School of Materials Science and Engineering, University of Science and Technology of China, 72 Wenhua Road, Shenyang, Liaoning 110016, China.
[5]Australian Nuclear Science and Technology Organisation, Locked Bag 2001, Kirrawee DC NSW 2232, Australia.

\# These authors equally contributed to the work.
\* Corresponding authors: tongx@ihep.ac.cn, bingli@imr.ac.cn

## Abstract:

Solid-state refrigeration based on caloric effects is an energetically efficient and environmentally friendly technology, which is deemed as a potential alternative to the conventional vapor-compression technology. One of the greatest obstacles to the real application is the huge driving fields. Here, we report a giant barocaloric effect in inorganic $NH_4I$ with maximum entropy changes of $\Delta S_{\text{BCE}}^{\max}$ ~89 J $K^{-1}$ $kg^{-1}$ around room temperature, associated with the orientationally order-disorder phase transition. The phase transition temperature, $T_t$, varies dramatically with pressure in a rate of $dT_t/dP$ ~0.81 K $MPa^{-1}$, which leads to a very much small saturation driving pressure of $\Delta P$ ~20 MPa, an unprecedentedly large caloric strength of $|\Delta S_{\text{BCE}}^{\max}/\Delta P|$ ~4.45 J $K^{-1}$ $kg^{-1}$ $MPa^{-1}$, as well as a broad temperature window of ~68 K under an 80 MPa driving pressure. Comprehensive characterization of the crystal structure and dynamics by neutron scattering measurements reveals a strong reorientation-vibration coupling that is responsible for the large pressure sensitivity of $T_t$. This work is expected to advance the practical application of barocaloric refrigeration.

To tackle climate change and realize the United Nations' Sustainable Development Goals, the first priority should be given to the decarbonization of heating and cooling sectors[1]. Nowadays, the vapor-compression technology is extensively employed for civil and industry refrigeration, which leads to two serious environmental concerns. On the one hand, billions of running fridges, air conditioning and heat-pump units are swallowing ~25% to ~30% of the electricity, and this demand is expected to continuously grow by several times in coming decades[2-5]. On the other hand, currently used refrigerants have thousand-time stronger global warming potential compared to $CO_2$[6]. For example, the global warming potential of the popular R134a is about 1300. Therefore, it is urgent to establish a low-carbon refrigeration solution.

Within such a context, the solid-state refrigeration technology based on caloric effects becomes a promising alternative. Caloric effects usually include magnetocaloric[7], electrocaloric[8], elastocaloric[9] and barocaloric effects[10], which characterize the thermal effects during a solid-state phase transition induced by a specific external field such as magnetic field, electric field, stress and pressure, respectively. In the entire refrigeration process, the working material stays solid and thus this technology is emission-free. As far as the energy efficiency is concerned, cooling systems working with caloric materials are considerably competitive as expected to reach 60–70% of the Carnot limit[11].

However, one of the greatest obstacles to the large-scale application of the caloric cooling technology is the difficulty that large caloric effects can be only achieved under huge driving fields in current leading materials. For instance, the magnetic fields used to stimulate metamagnetic or magneto-structural transitions in magnetocaloric materials are generally larger than 2 T, which requires heavy and expensive rare-earth-based permanent magnets or superconducting magnets[12]. With respect to the electrocaloric materials, the electric fields are in the magnitude of kV m$^{-1}$ or even MV m$^{-1}$, which might create breakdown phenomena and hence influence the operation reliability and cycling lifetime[13]. In the case of leading elastocaloric materials, the typical driving stress is as large as 700 MPa to obtain good refrigeration performances[14]. As for barocaloric materials, the required pressure is usually above 200 MPa for most intermetallics[15, 16] and it is reduced down to about 100 MPa in the recently discovered plastic crystals[10]. Nonetheless, the development of excellent caloric materials with a smaller driving field remains highly challenging.

In this paper, we report a giant barocaloric effect around room temperature in commercially available ammonium iodide ($NH_4I$) compound. The phase transition temperature in $NH_4I$ displays a very high sensitivity to driving pressure, which renders $NH_4I$ a very small saturation driving pressure and makes it one of most efficient and cost-effective caloric materials as estimated by barocaloric strength (maximum isothermal entropy change divided by driving force). In addition, thorough studies on the microscopic crystal structures and dynamics using neutron scattering techniques demonstrate that the excellent barocaloric effect is mainly attributed to the configuration entropy changes of $[NH_4]^+$ tetrahedra in the frameworks formed by $I^-$ ions and that the large sensitivity of phase transition temperature to external pressure is linked to the strong coupling between molecular reorientation and lattice vibration.

## Results

**Barocaloric effects in NH$_4$I.** The calorimetric behaviors and the barocaloric effects in NH$_4$I are studied with the differential scanning calorimetry measurements over the temperature range of 230 K to 340 K, under several constant external pressures (**Methods** and **Supplementary Fig. 1**). Following the heat flow data (**Supplementary Fig. 2a**), a phase diagram is established. As shown in **Fig. 1a**, a sharp phase transition is observed at ~243 K on cooling or ~268 K on heating under ambient pressure. A large thermal hysteresis of $\Delta T_{\text{hys}}$ ~25 K indicates the first-order nature of this phase transition. This phase transition happens between the intermediate-$T$ $\beta$-phase (space group $Pm\bar{3}m$) and the high-$T$ $\alpha$-phase ($Fm\bar{3}m$)[17, 18], which will be confirmed by the neutron scattering studies in following sections. In addition, it is found that the phase transition temperature varies strongly with external pressure, being consist with the previous pressure-dependent Raman and thermal conductivity studies[17, 19]. According to **Fig. 1a**, the d$T_t$/d$P$ is as large as ~0.81 K MPa$^{-1}$ on cooling and ~0.79 K MPa$^{-1}$ on heating, which are much larger than those of other leading barocaloric materials as summarized in **Fig. 1c**[3, 5, 10, 15, 16, 20-34]. This large d$T_t$/d$P$ value means that application of an 80 MPa driving pressure could open up a ~68 K working temperature window (on cooling).

The isobaric entropy with respect to 230 K, $S'(T,P) = S(T,P) - S(230\text{ K}, P)$, could be derived from the heat flow (Supplementary **Fig. 2b**), which are then used to assess the isothermal barocaloric entropy change, $\Delta S_{\text{BCE}}(T, \Delta P)$, by $\Delta S_{\text{BCE}}(T, \Delta P) = S'(T,P) - S'(T,P')$ [10, 24]. As shown in **Fig. 1b**, it is found that the maximum value of $\Delta S_{\text{BCE}}^{\max}(T, \Delta P)$ on cooling is ~89 J K$^{-1}$ kg$^{-1}$ (~71 J K$^{-1}$ kg$^{-1}$ on heating). This value is comparable with other state-of-the-art barocaloric materials (Supplementary **Fig. 2c**)[3, 5, 10, 15, 16, 20-34]. Noticeably, the giant d$T_t$/d$P$ value leads to a very small saturation driving pressure of 20 MPa (see solid line in **Fig. 1c**), hence a large barocaloric strength, defined by $|\Delta S_{\text{BCE}}^{\max}/\Delta P|$, of ~4.45 J K$^{-1}$ kg$^{-1}$ MPa$^{-1}$. As summarized in **Fig. 1d**, the $|\Delta S_{\text{BCE}}^{\max}/\Delta P|$ value for NH$_4$I is also much larger than most other barocaloric materials, making it one of the most high-efficient solid-state refrigeration materials, especially compared with the inorganic group.

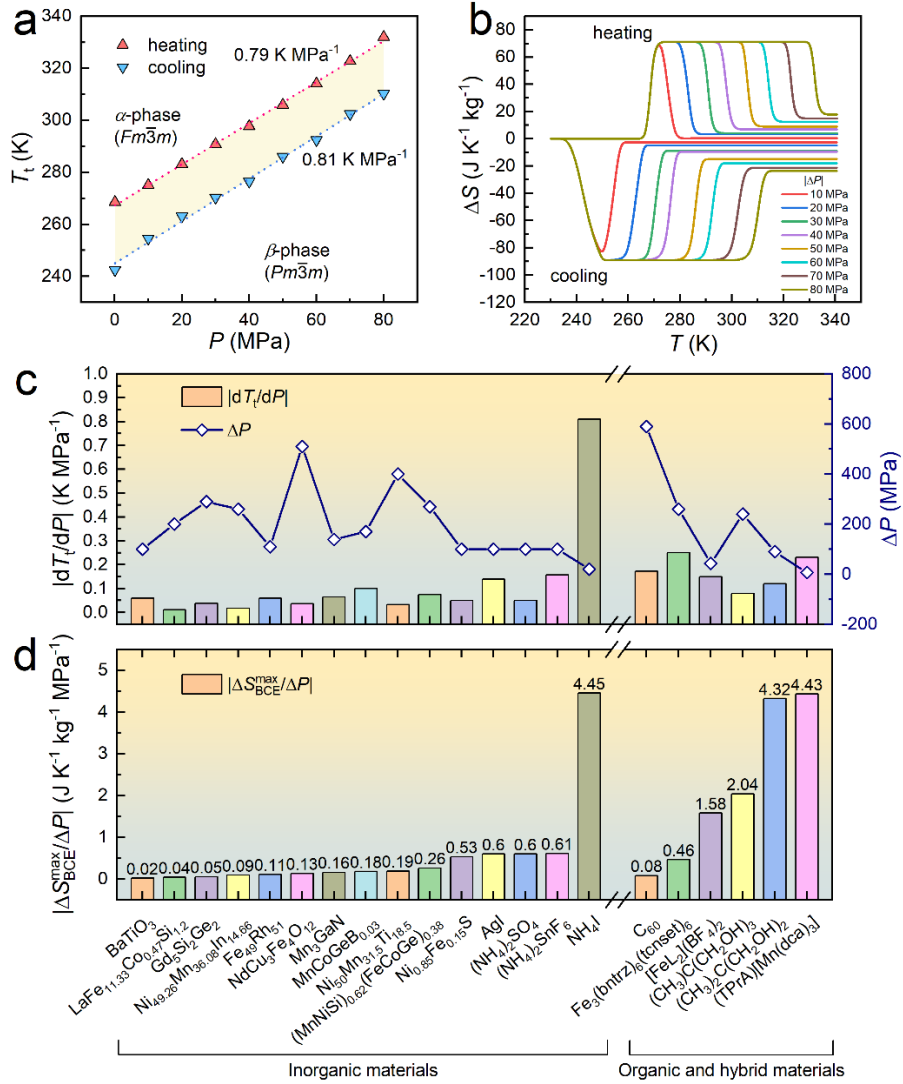

**Fig. 1 Barocaloric effects of NH$_4$I. a**, Phase diagram of NH$_4$I as functions of temperature and pressure. The cooling (down-triangle) and heating transition (up-triangle) temperatures are derived from calorimetric data of **Supplementary Fig. 2a**. **b**, Isothermal entropy change, Δ$S$, for the cooling and heating processes. **c**, Much larger pressure-dependent transition temperature variation |d$T_t$/d$P$| and smaller saturation driving pressure compared with other giant barocaloric materials[3, 5, 10, 15, 16, 20-34]. **d**, High barocaloric efficiency as estimated by the barocaloric strength of $|\Delta S_{\text{BCE}}^{\max}/\Delta P|$. All details are summarized in **Supplementary Table 1**.

**Phase transitions as a function of temperature.** Details about the phase transition behaviors in NH$_4$I are studied with inelastic neutron scattering (INS) measurements using the Time-of-Flight Spectrometer, PELICAN, at the Australian Centre for Neutron Scattering (see **Methods**)[35]. **Fig. 2a** illustrates three INS spectra $S(Q,\omega)$ as a function of energy and momentum transfer, collected at three representative temperature points, 160 K, 260 K and 300 K, respectively. These spectra exhibit different features, corresponding to three different phases in the NH$_4$I compound. Then, the elastic section $S(Q)$ is extracted to obtain the diffraction patterns by integrating $S(Q,\omega)$ over [-0.3, 0.3] meV. The results are shown in **Figs. 2b,c**. An obvious phase transition can be found at ~275 K, corresponding to the transition from intermediate-$T$ $β$-phase with a space group of $Pm\bar{3}m$ to the high-$T$ $α$-phase with $Fm\bar{3}m$ [18, 36] (schematic crystal structures are shown in **Supplementary Fig. 3**). However, it is quite difficult to discern another phase transition between the low-$T$ $γ$-phase ($P4/nmm$) and the intermediate-$T$ $β$-phase, as the $γ$-phase derives from a tiny distortion from the $β$-phase[18, 37].

In order to accurately track the phase transition, the atomic mean-squared-displacement (MSD) are analyzed with the Debye-Waller factor fitting of the elastic structure factor, $S(Q)$ (see **Supplementary Fig. 4**)[38]. The obtained MSD as a function of temperature is shown in **Fig. 2d**. The MSD across the transition of $β→α$ exhibits an abrupt jump. In addition, it is noted that the temperature dependences of MSD present a crossover at ~193 K, which corresponds to the phase transition of $γ→β$. Given the much larger cross section of H (80.26 barn) than those of I (0.31 barn) or N (0.5 barn), the obtained MSD in **Fig. 2d** mainly reflects the thermal fluctuation behaviors of H atoms. In fact, the sliced $S(Q,\omega)$ curves over the $Q$ range of [1.55, 1.65] Å$^{-1}$ also show strong broadening above 193 K, and this broadening develops continuously with increasing temperature until 280 K as shown in **Fig. 2e**. These two temperature points correspond exactly to the phase transitions of $γ→β$ and $β→α$.

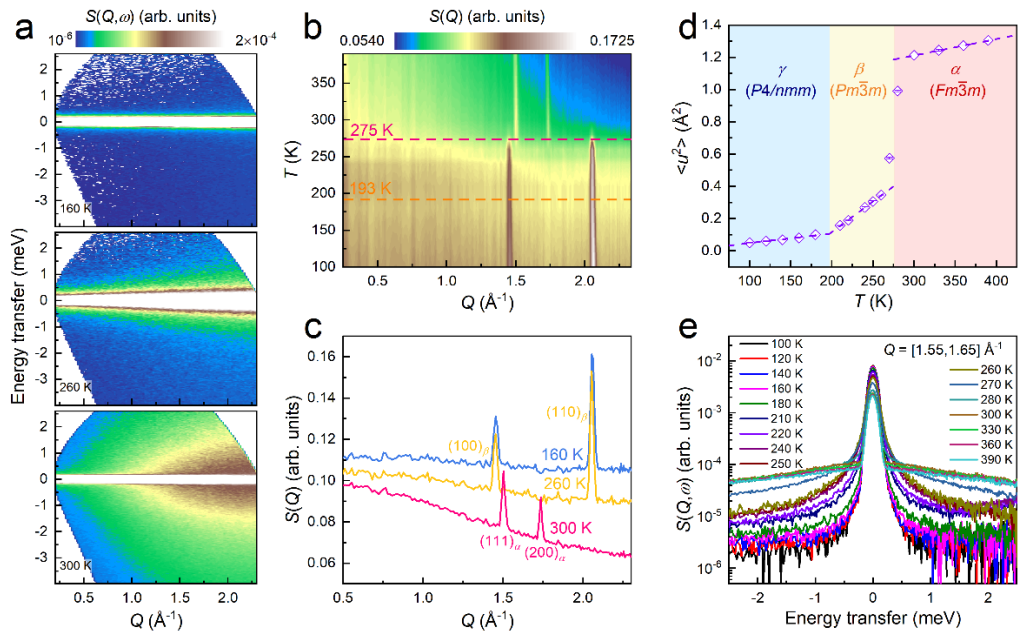

**Fig. 2 Phase transitions as function of temperature in NH$_4$I. a**, Dynamic structure factor, $S(Q,\omega)$, for NH$_4$I at 160 K, 260 K and 300 K measured at PELICAN with $E_i$ = 3.72 meV. **b**, The contour image of elastic structure factor $S(Q)$ as function of temperature. **c**, Comparison the $S(Q)$ at temperature of 160 K, 260 K and 300 K. The Bragg peaks for the high-$T$ $α$-phase and the intermediate-$T$ $β$-phase are marked with subscripted '$α$' and '$β$', respectively. **d**, Experimental mean squared displacement, $<u^2>$, determined from the fitting of the elastic window of the $S(Q,\omega)$ to the Debye-Waller effect, $S(Q) \propto \exp(-Q^2\langle u^2\rangle/3)$, see **Supplementary Fig. 4**. **e**, Plot of sliced $S(Q,\omega)$ over the $Q$ range of [1.55, 1.65] Å$^{-1}$ as a function of energy.

**Order-disorder transition and reorientational dynamics.** The broadenings centered around 0 meV as observed in **Figs. 2a,e** are signals of quasi-elastic neutron scattering (QENS). QENS spectra could be used to study the dynamics of molecular rotations in materials[39, 40]. The $S(Q,\omega)$ over the $Q$ range of [1.55, 1.65] Å$^{-1}$ below ~190 K could be fitted with a delta function convoluted with the instrumental resolution function plus a linear background (**Fig. 3a**). This implies that hydrogen atoms in [NH$_4$]$^+$ tetrahedra stay in the lattice of the low-$T$ $\gamma$-phase, without any jump or rotation, in agreement with the crystallographic analysis based on diffraction patterns[18]. On the other hand, a good fitting for the spectra above 190 K needs one more Lorentzian component as depicted in **Figs. 3b,c**, which is indicative of reorientational or jumping behaviors in the $\beta$- and $\alpha$-phases.

The relaxation time, $\tau$, of the reorientation modes could be estimated by the linewidths ($\Gamma$) of the Lorentzian profile with the formula of $\tau = 2\hbar/\Gamma$ [40]. As shown in the inset of **Fig. 3d**, the average $\Gamma$ at 300 K is ~3.55 meV ($\tau$ ~0.37 ps), which is ~20 times smaller than the value of ~0.17 meV ($\tau$ ~7.6 ps) for the 260 K spectrum. Therefore, the reorientation mode in the high-$T$ $\alpha$-phase is ~20 times faster than that in the intermediate-$T$ $\beta$-phase, in agreement with the previous reports [41]. It is also noted that $\Gamma$ for both 260 K and 300 K spectra are almost independent from $Q$. Such behaviors demonstrate localized nature of the reorientation modes, similar to what happen in perovskite CH$_3$NH$_3$PbI$_3$[42] and nano-NaAlH$_4$[42, 43]. In addition, the activation energies of the reorientation modes are examined by fitting the temperature dependent $\Gamma(T)$ to the Arrhenius relation. The activation energy for the mode in the intermediate-$T$ $\beta$-phase is as high as 117(1) meV, while the reorientational motion in the high-$T$ $\alpha$-phase only needs to overcome a much smaller energy barrier of 18(2) meV (**Fig. 3d**).

The ratio between the elastic intensity (integrated area below the delta function) and the total intensity (elastic intensity plus the QENS intensity below the Lorentzian profile) gives rise to elastic incoherent structure factor (EISF, see **Methods**). The $Q$ dependence of EISF could be used to analyze the reorientational geometry. Here, several models are employed to reproduce the experimentally determined EISF, including twofold ($C_2$) and/or threefold ($C_3$) jumps, cubic tumbling as well as isotropic rotational diffusion (see **Methods** or refer to [44]). As shown in **Fig. 3e**, the EISF at 260 K can be well reproduced by the model of cubic tumbling jump. In this model, hydrogen atoms in [NH$_4$]$^+$ tetrahedra could jump among the eight corners of a cube. The eight corners of this cube localize on the lines between the [NH$_4$]$^+$ and the eight I$^-$ nearest-neighbors as illustrated in the inset of **Fig. 3e**. In this configuration with $T_d$ point group, the [NH$_4$]$^+$ tetrahedra have 2 reorientational freedoms[45]. In addition, the collinear arrangement of the N-H⋯I bond could form an energy minimum, which leads to a large hindering barrier to the rotational motion in the intermediate-$T$ $\beta$-phase. This is confirmed by the large activation energy of 117(1) meV as determined in **Fig. 3d**.

With respect to the high-$T$ $\alpha$-phase, things become more intricate. In this case, each [NH$_4$]$^+$ ion is surrounded by six I$^-$ anions, which form an octahedral cage with $O_h$ point group symmetry. The four tetrahedrally arranged H atoms suffer a strong geometric frustration, and cannot achieve a close approach to the octahedrally distributed I$^-$ anions simultaneously[46]. To obtain clear picture about the hydrogen distribution in the high-$T$ $\alpha$-phase, a lot of models have been proposed in early literatures[18, 46-48], such as single approach, double approach and triple approach models[46] or isotropic model[48]. However, the previous QENS analysis, reported by Goyal *et al*. in 1979, indicated that none of these models could give an appropriate description about the reorientational motion of hydrogen atoms in [NH$_4$]$^+$ tetrahedra[41]. Nonetheless, the experimental EISF curve obtained at 300 K in this work points to an isotropic rotation model in the high-*T* *α*-phase as shown in **Fig. 3f**, which is in agreement with the nuclear magnetic resonance studies[48].

Overall, either the single approach, double approach and triple approach models as proposed by Levy *et al.*[46], or the isotropic rotation model as illustrated in this work are all energetically equal, giving larger degenerated states and hence larger freedom of disorders in the high-*T* *α*-phase than what happen in the intermediate-*T* *β*-phase[46, 49]. Even with the single approach model, as shown in the inset of **Fig. 3f**, six-fold configurations are allowed (without considering any possible rotation about the N-H bond, see **Supplementary Fig. 3**). This leads to a configuration entropy change of 63 J K$^{-1}$ kg$^{-1}$ (see **Methods**), comparable to the experimentally obtained entropy change across the *β*→*α* transition as shown in **Fig. 1b**.

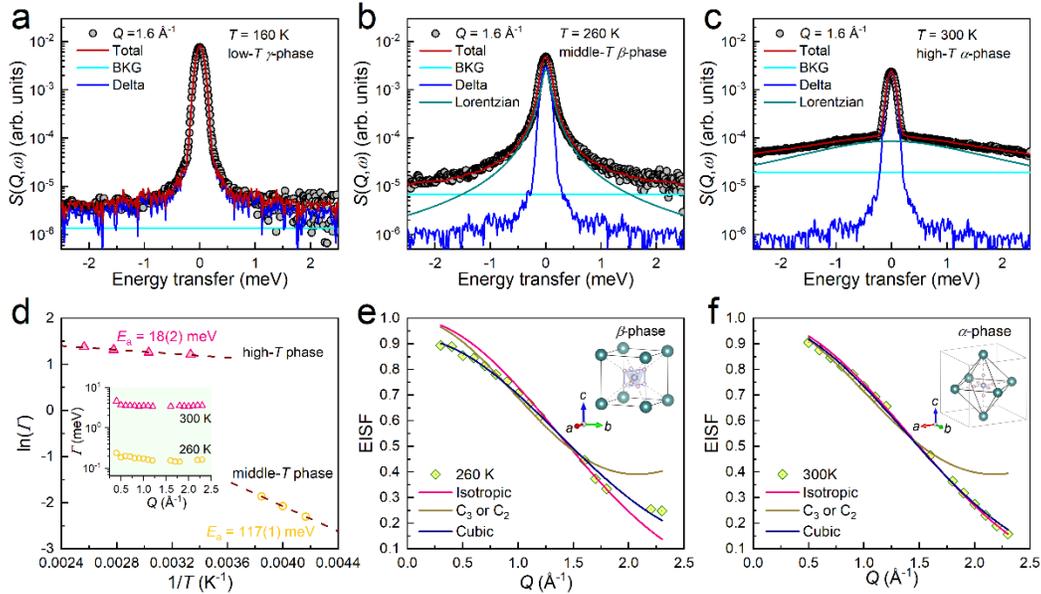

**Fig. 3 Reorientational dynamics of ammonia molecule. a-c**, Spectral fitting of the sliced *S*(*Q*,*ω*) over a *Q* range of [1.55, 1.65] Å$^{-1}$ at (**a**) 160 K, (**b**) 260 K and (**c**) 300 K, respectively. One delta function with a constant background (BKG) could make a good fitting for the spectrum at 160 K, while one more Lorentzian profile is needed at 260 K and 300 K. **d**, Temperature dependence of the full width at half maximum, $\Gamma$, of the Lorentzian components for the intermediate-*T* *β*-phase and the high-*T* *α*-phase, fitted to the Arrhenius equation, $\Gamma \propto \exp\left(-\frac{E_a}{k_B T}\right)$, where $E_a$ is the activation energy for jumping reorientations and $k_B$ is the Boltzmann constant. Inset shows *Q* dependences of the $\Gamma$. **e**,**f**, Experimental EISF compared with few curves associated with different jumping reorientation models at 260 K and 300 K. Insets shows the schematic figures of the I$^-$ nearest-neighbors surrounding the [NH$_4$]$^+$ ions in the *β*- and *α*-phases. Following the symmetric operations, the H atoms could dwell on any of the diagonal line in the *β*-phases, while the H atoms residing along the [100] direction would give a sixfold possibilities of steric distribution in the *α*-phase (single approach model, see **Supplementary Fig. 3**).

**Strong reorientation-vibration coupling.** The neutron weighted phonon density of states (DOSs) were measured with the NH$_4$I powder sample on the PELICAN at ACNS (**Methods**), and the results are shown in **Fig. 4**. The phonon DOS pattern at 160 K contains six well-defined peaks over the energy transfer range of 0-80 meV. According to the previous INS studies, the ~4.8 meV and ~7 meV peaks are the acoustic (marked as 'A') phonon bands, and the ~18 meV peak is the optical (marked as 'O') phonon band[50, 51]. However, the other three optical phonon bands at higher energy range are not reported in the Refs. of 50, 51. But the corresponding counterparts can be observed in NH$_4$Br[52], although the peak positions exhibit some differences because of the different molecular masses or chemical bonding strengths in NH$_4$I and NH$_4$Br. As referred to NH$_4$Br, these three high-energy optical phonon bands come from the libration motions of the hydrogen atoms in [NH$_4$]$^+$ tetrahedra.

One obvious phenomenon about the phonon DOSs is the quick broadening of these six peaks with increasing temperature, which then become featureless above the $\beta \rightarrow \alpha$ transition temperature of ~275 K. In addition, it is observed that the optical phonon bands, especially the O$_{II}$, O$_{III}$ and O$_{IV}$ ones, show dramatic softening with increasing temperature as delineated by the dashed lines in **Fig. 4**. These two features of the phonon DOSs demonstrate that the lattice vibration potentials of NH$_4$I exhibit strong anharmonicity. In combination with the fact that the hydrogen atoms in [NH$_4$]$^+$ tetrahedra exhibit different reorientational dynamics with the development of phonon anharmonicity (**Fig. 3**), it can be concluded that NH$_4$I compound has strong coupling between molecular reorientations and lattice vibrations, and that its phase transitions strongly correlate to the reorientational dynamics of the [NH$_4$]$^+$ tetrahedra. This reorientation-vibration coupling dominated phase transitions in NH$_4$I can also be understood from the viewpoint of thermodynamics. With increasing temperature, the rising rotational entropy from the [NH$_4$]$^+$ tetrahedra would continuously lower the Gibbs free energy of the high-$T$ phases until the phase transition occurs, similar to the entropy-driven structural transition in formamidinium lead iodide perovskite[53].

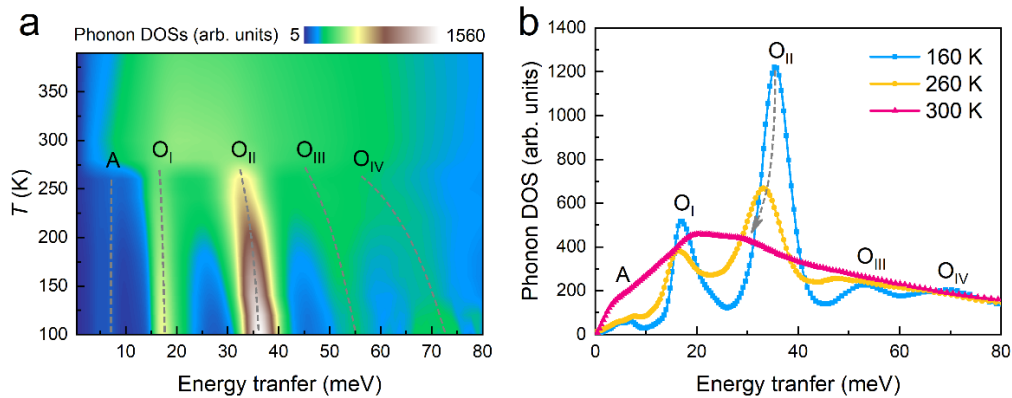

**Fig. 4 Variation of lattice dynamics with temperature. a**, Contour plot of neutron-weighted phonon density of states (DOSs) as function of temperature, measured at PELICAN. **b**, Comparison of phonon DOSs curves at 160 K, 260 K and 300 K, respectively. The dashed lines are used for guiding eyes to track the variation of the phonon bands with temperature. 'A' and 'O' denote the acoustic and optical phonons, respectively.

**Responses of dynamic behaviors to external pressure.** The responses of the phase transition to external pressure are also studied using INS with different pressures. As shown in **Figs. 5a,b**, the QENS broadening of the 0.1 MPa data at 300 K is quite broader than what observed in the 300 MPa data at the same temperature. This comparison become clearer in the sliced curves over the $Q$ range of [1.55, 1.65] Å$^{-1}$ as shown in **Figs. 5c,d**. It is obvious that the faster (larger $\Gamma$) reorientational mode with larger orientational degrees of freedom is suppressed to the slower (smaller $\Gamma$) mode with smaller degrees of freedom. In addition, with the suppression of the faster reorientational mode, two peaks emerge at ~19 meV and ~34 meV from the featureless phonon DOS of the high-$T$ $\alpha$-phase, corresponding to the O$_I$ and O$_{II}$ bands in the intermediate-$T$ $\beta$-phase as shown in **Figs. 5e,f**. Overall, application of external pressure induces changes in both the reorientational and lattice dynamics, and this provides a direct description about the underlying microscopic mechanism for the pressure-driven colossal barocaloric effect in NH$_4$I.

It is worth to go back to the issue about the reorientation-vibration coupling in NH$_4$I. The simultaneous variations of crystallographic symmetries, anion coordination, mean squared displacement, lattice dynamics as well as the reorientational dynamics imply that the N-H⋯I hydrogen bonds between [NH$_4$]$^+$ and I$^-$ play the key roles in the coupling between the molecular reorientations and lattice vibrations, as also suggested in other literatures[17, 54, 55]. In fact, the strong optical phonon softening as observed in **Fig. 4** is believed to be a direct result of the continuous weakening of the hydrogen bonds with increasing temperature. With respect to application of external pressure at 300 K in **Fig. 5**, the hydrogen bonds could be enhanced dramatically with compression[17]. The intensified binding of the [NH$_4$]$^+$ tetrahedra to the lattice not only sets a higher barrier for the reorientational motion of [NH$_4$]$^+$ tetrahedra, but also imposes stronger restricting forces on the lattice. These restricting forces are large enough to induce a giant change in the coordination environments from six-octahedron configuration to eight-cube configuration. In turn, this transformation of coordination environments will extend the N-H⋯I bonds number, which will further enhance the reorientation-vibration coupling. This dramatically enhanced reorientation-vibration coupling across the $\alpha\rightarrow\beta$ phase transition manifests as the large increase of the reorientational activation energy from 18(2) meV in the high-$T$ $\alpha$-phase to 117(1) meV in the intermediate-$T$ $\beta$-phase. It is this large alteration of the reorientation-vibration coupling strength that gives rise to a huge pressure dependence of the $\alpha\rightarrow\beta$ phase transition temperature, and hence a small saturation driving pressure and a giant barocaloric strength of $|\Delta S_{BCE}^{max}/\Delta P|$ as shown in **Fig. 1**.

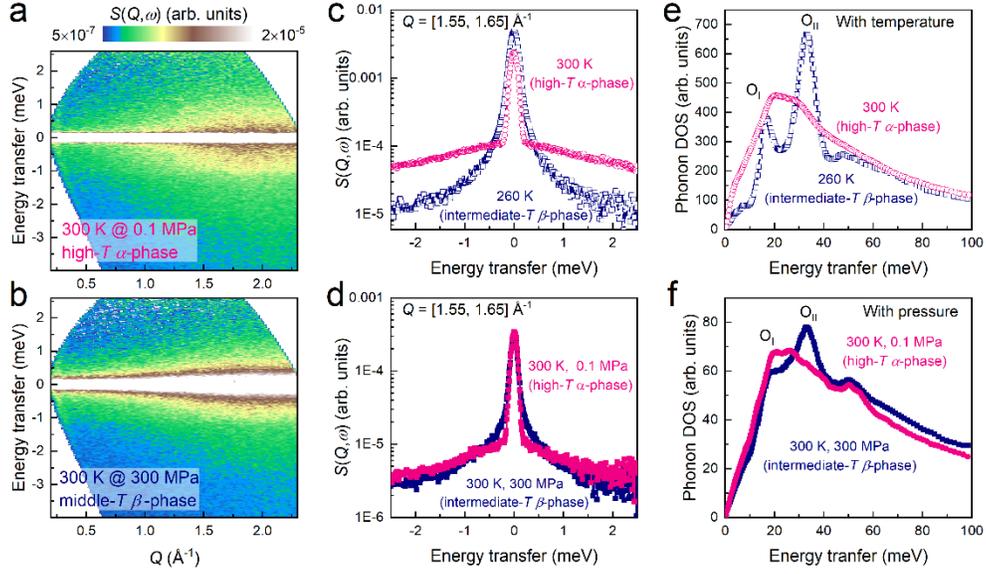

**Fig. 5 Responses of dynamic behaviors to external pressure. a,b**, The dynamic structure factor, $S(Q,\omega)$, for $NH_4I$ at 300 K with external pressure of (**a**) 0.1 MPa (ambient condition) and (**b**) 300 MPa, respectively. **c,d**, Comparison of the of $S(Q,\omega)$ sliced over $Q$ = [1.55, 1.65] Å$^{-1}$ between the high-$T$ phase and intermediate-$T$ phase in responses to (**c**) temperature and (**d**) external pressure. **e,f**, Comparison of the phonon DOSs between the high-$T$ phase and intermediate-$T$ phase in responses to (**e**) temperature and (**f**) external pressure.

## Conclusions

To summarize, we present a thorough study on the inorganic $NH_4I$ compound which exhibits a high-efficient barocaloric effect over a broad temperature range around room temperature. One of the most noticeable features of $NH_4I$ is the giant pressure sensitivity of the phase transition temperature. Although larger entropy change requires smaller $dT_t/dP$ as suggested by the Clausius-Clapeyron equation[28], larger $dT_t/dP$ value often bring more benefits in practice, including smaller and more accessible driving pressure and bigger barocaloric strength as demonstrated in $NH_4I$. The physics underlying the barocaloric effect in $NH_4I$ is analyzed using neutron scattering techniques. The high pressure-sensitivity of the phase transition temperature stems from the strong coupling between molecular reorientations and lattice vibrations, while the giant entropy change is largely contributed by the reorientational disorder of $[NH_4]^+$ tetrahedra. This work would inspire the discovery of giant barocaloric materials with high pressure-sensitive phase transition, and hence push a big step forward towards the realization of high-efficient and sustainable barocaloric refrigeration.

## Methods:

**Sample preparation**: The $NH_4I$ powder sample with 99.999% purity was purchased from Aladdin. The powder sample was checked with XRD at room temperature (see **Supplementary Fig. 1**).

**Caloric measurements**:

The heat flow measurements were performed using a high-pressure differential scanning calorimeter, μDSC, Setaram. ~20 mg $NH_4I$ powder sample was sealed in a high-pressure vessel made of Hastelloy, while an empty vessel was also used as a reference during the measurement. Desirable hydrostatic pressures were generated and maintained by controlling argon gas pressure through the high-pressure gas panel. The measurements were carried out over the temperature range of 230 K to 340 K under constant pressure of 0.1, 10, 20, 30, 40, 50, 60, 70, and 80 MPa, and a ramping rate of 1 K min$^{-1}$ was used for both the cooling and heating processes.

**INS/QENS measurements.** The INS and/or QENS experiments were conducted using the cold-neutron time-of-flight spectrometer, PELICAN, at the Australian Centre for Neutron Scattering, ANSTO[35]. The instrument was configured with an incident neutron wavelength of 4.69 Å, affording an incident energy of 3.72 meV with an energy resolution of 0.135 meV at the elastic line. The $NH_4I$ powder sample was sealed in an annular aluminum can. The measurements were carried out from 100 K to 390 K to cover the two continuous phase transitions in the sample. The empty can was measured in the same conditions for background subtraction. In addition, a standard vanadium sample was also measured for detector normalization and to determine the energy resolution function. The high-pressure neutron scattering measurements were performed with the same configuration at 0.1 MPa and 300 MPa at 300 K. A bespoke high-pressure cell made of Be-Cu alloys was used, and KBr powder sample was employed to calibrate the actual pressures. The data reduction, including background subtraction and detector normalization were performed using the Large Array Manipulation Program (LAMP)[56], while the sliced QENS spectra were analyzed with the Pan module built-in the Data Analysis and Visualization Environment (DAVE)[57].

**Analysis of the rotational dynamics**. In the INS spectrum, the broadening underneath the elastic peak is signals of QENS, which is associated with diffusive and/or reorientational motions. Assuming one single reorientation mode, the QENS dynamics structural factor $S(Q,\omega)$ can be modeled as[40, 58]:

$$S(Q,\omega) = f \times [A_0(Q)\delta(\omega) + A_1(Q)L(Q,\omega)] \otimes R(Q,\omega) + b(Q,\omega)$$

where $\delta(\omega)$ is a delta function representing the elastic peak at the zero energy transfer, $A_0(Q)$ and $A_1(Q)$ are the weights of the elastic and quasielastic scattering, $f$ is scale factor, $R(Q,\omega)$ is experimentally determined resolution function, and $b(Q,\omega)$ is the linear background. The symbol $\otimes$ describes numerical convolution between the elastic/quasielastic components and instrumental resolution. The quasielastically broadened energy distribution could be well depicted by Lorentzian function:

$$L(Q,\omega) = \frac{1}{\pi}\frac{\Gamma(Q)}{(\hbar\omega)^2 + \Gamma(Q)^2}$$

where $\Gamma(Q)$ is the half width at half maximum, corresponding to the jump frequency of the hydrogen atoms in [NH$_4$]$^+$ tetrahedra between different orientations.

To analyze the reorientational geometry of the [NH$_4$]$^+$ tetraheda, the elastic incoherent structure factor (EISF) was extracted from fitting of the QENS spectra:

$$\text{EISF} = \frac{A_0(Q)}{A_0(Q) + A_1(Q)}$$

Three models were considered to reproduce the experimentally determined EISF[44]:

$C_2$ and/or $C_3$ jumps:

$$\text{EISF} = \frac{1}{2}[1 + j_0(Qd)]$$

Cubic tumbling:

$$\text{EISF} = \frac{1}{8}\left[1 + 3j_0\left(\frac{2}{\sqrt{3}}Qd\right) + 3j_0(Qd) + j_0(2Qd)\right]$$

Isotropic rotational diffusion:

$$\text{EISF} = j_0^2(Qd)$$

where $d$ is rotational radius and $j_0(x) = \frac{sin(x)}{x}$ is the spherical Bessel function of zeroth order[40, 44].

When the possible orientational geometries below and above a phase transition are determined, the configurational entropies across this phase transition can be estimated by the formula as following[59]:

$$\Delta S_{\text{conf}} = R \ln \frac{N_2}{N_1}$$

where $R$ is the gas constant, $N_1$ and $N_2$ are the numbers of possible orientational configurations below and above a phase transition temperature.

**Analysis of the lattice dynamics.** In the incoherent one-phonon approximation, the measured dynamic structure factor with powder sample correlates to the phonon density of states (PDOSs) $g(E)$[60, 61]:

$$g(E) = A \times \left\langle 4M \frac{\exp(2W)}{\hbar^2 Q^2} \frac{E}{n(E,T) + \frac{1}{2} \pm \frac{1}{2}} S(Q, \omega) \right\rangle$$

where $A$ is scale factor, $M$ and $\exp(-2W)$ is the mass and Debye-Waller factor, while the + and − signs denote energy loss or energy gain of neutrons, respectively. $n(E,T)$ is the Bose-Einstein occupation factor, defined as $n(E,T) = [\exp(E/k_B T) - 1]^{-1}$, where $k_B$ is the Boltzmann constant. The brackets $\langle \cdots \rangle$ represent the average operator over all $Q$ range at a given energy. In a polyatomic material, the experimental determined PDOSs correspond to neutron-weighted phonon DOS, $g_{\text{NW}}(E)$, because different elements have different scattering cross sections $\sigma$ and atomic masses $M$:

$$g_{\text{NW}}(E) = \sum_i f_i \frac{\sigma_i}{M_i} g_i(E)\exp(-2W_i)$$

here $i$ represents different elements, $f_i$ is the atomic concentration, $g_i(E)$ is the real partial phonon DOS of the element $i$. In this work, due to the limited maximum value (~3.72 meV) in the energy loss side, only the energy gain side of the $S(Q,\omega)$ up to 80 meV was used to extract the $g_{\text{NW}}(E)$. In addition, it is noted that the values of $\sigma/M$ for N, H and I are 0.82, 82.02 and 0.03 barn/amu, respectively. Therefore, the obtained $g_{\text{NW}}(E)$ mainly reflects the vibrational information of hydrogen atoms in the lattices.

## Data availability

The data that support the findings of this study are available from the corresponding author upon request.

## Acknowledgements

We acknowledge the support by the National Key Research and Development Program of China (Grant Nos. 2020YFA0406001 and 2020YFA0406002), the Key Research Program of Frontier Sciences of Chinese Academy of Sciences (Grant No. ZDBS-LY-JSC002), the Liaoning Revitalization Talents Program (Grant No. XLYC1807122), National Natural Science Foundation of China (Grant Nos. 11804346 and 11875265), the Scientific Instrument Developing Project of the Chinese Academy of Sciences (Grant No. 284(2018)), Guangdong Basic and Applied Basic Research Foundation (No. 2019B1515120079), R & D projects in key areas of Guangdong Province (No. 2019B010941002), Dongguan Introduction Program of Leading Innovative and Entrepreneurial Talents (No. 20191122), and Open Funding Project of the Laboratory of Artificial structures and Quantum Control, Ministry of Education, at Shanghai Jiao Tong University (no. 2020-05). We acknowledge the beam time granted by ANSTO (Proposal no. 8268 and 8318).

## Author contributions

B.L. proposed the project. J.Q. carried out thermal characterizations of the samples with help of W.J.R. and B.L. Neutron scattering measurement with temperature and pressure were performed by D.H.Y., Q.Y.R., J.Q., W.L.S. and B.L. The high-pressure cells were prepared by B.Y., Q.Y.R and X.T. Analysis of thermal properties and neutron scattering spectra were conducted by Q.Y.R. with the help from J.Q., W.L.S., T.H.W., Z.D.Z., X.T. and B.L. Q.Y.R. drafted the manuscript. All authors edited and finalized the manuscript.

## Competing interests

The authors declare no competing interests.

## Additional information

Supplementary information is available for this paper at XXX.